%% file: techreport.tex
\documentclass[reprint,aps,prl,nobalancelastpage]{revtex4-1}
\usepackage{graphicx}
\usepackage{placeins}

\begin{document}
\title{Technical Report: Performance of the \emph{expected force} on AS-level Inernet topologies.}
\author{Glenn Lawyer}
\affiliation{Max Planck Institute for Informatics}
\date{11 March 2014, TR-exf-nt-001}
\maketitle

The \emph{expected force} (ExF) is a metric which quantifies the spreading power of all network nodes. It is derived from a continuous-time epidemiological perspective, and uses the combinatorics inherent in local topology to compute the influence of each node. This is in direct contrast to the usual approach to measuring centrality, which is counting some type of walk on a network \cite{Estrada2010,Benzi2013}. 
The ExF has been previously shown to strongly and significantly outperform other existing centrality measures in predicting the outcome of spreading processes on many complex networks~\cite{Lawyer2014}.

Infrastructure networks, however, are different from other networks in that they are strongly constrained by specific engineering and economic constraints. 
A study of Internet connectivity on the router-level found that the high performance topologies which result from a design process are extremely rare to occur by chance, whereas more likely random networks have poor performance~\cite{Li2004}. Likewise, airline traffic networks maximize dynamic traffic flows, 
whereas social networks often experience bottlenecks to dynamic transfer of information~\cite{Pan2011}.
This raises the question of how the ExF will perform  on highly engineered network topologies.

This technical report presents an evaluation of the ExF on real world snapshots of the Internet's autonomous system (AS) level level connectivity. As in~\cite{Lawyer2014}, comparison is made to the eigenvalue centrality and the k-shell. 
The ExF is shown to be strongly predictive of node influence, significantly outperforming the other measures.

\section{Methods}
Five daily snapshots of the Internet's AS-level topology were downloaded from UCLA's \emph{Internet AS-level Topology Archive}
\footnote{http://irl.cs.ucla.edu/topology/}.
These snapshots are inferred from observations taken from 134 BGP data collectors.
Snapshots from 2014-03-03, 2014-02-01, 2014-01-01, and 2013-1201 capture current trends and topologies, while the snapshot from 2010-10-08 allows a touch of historical perspective.
The networks are presented in Table 1.

Analysis of the network degree distributions showed large proportion of nodes with degree one which connected to more central nodes. This suggests that the modified version of the ExF would be the appropriate measure for SIS/SIR processes. 
The modification is motivated in that under such network topologies and processes, a node's chance of realizing its ExF depends on a successful first transmission. To reflect this, the standard ExF is adjusted by the (scaled) log of the degree of the seed node~\cite{Lawyer2014}.
Preliminary results confirmed that this adjustment was helpful on AS level maps of internet connectivity.

One thousand nodes were selected uniformly at random from each network. The ExF (or ExF$^M$), the eigenvalue centrality, and the k-shell of each node were measured.
One hundred SI, SIS, and SIR spreading processes were seeded from each node. 
For each network, the correlation between node metrics and mean outcomes of the spreading processes was computed.

\begin{figure}[h]
\includegraphics[width=\columnwidth]{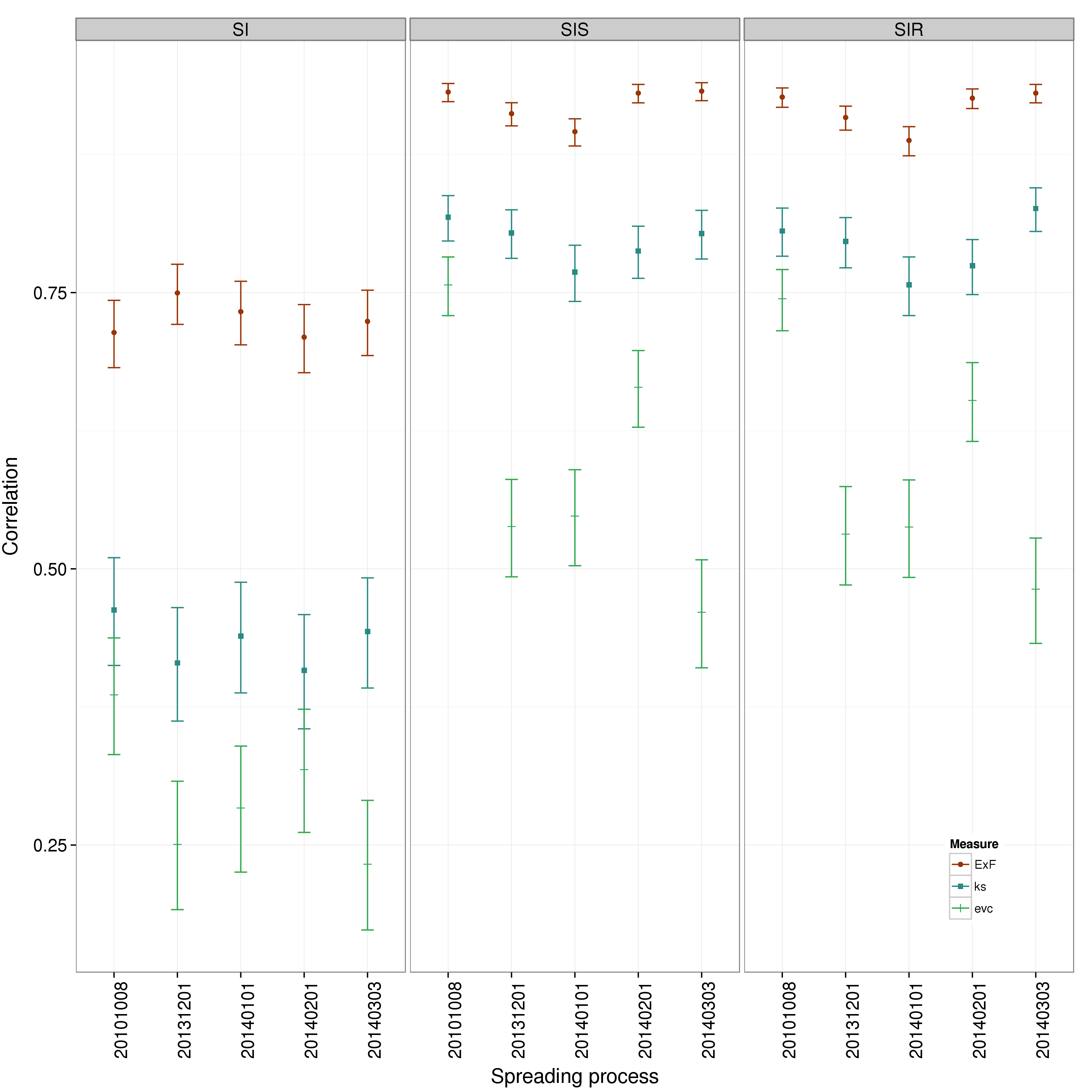}
\caption{The ExF has higher correlation and tighter confidence bounds than the other measures on all networks and spreading processes tested.
ExF= expected force, ks=k-shell, evc= eigenvalue centrality. The values used to create the plot are given in tabular form in the supplementary tables.}
\end{figure}

\begin{table}[ht]
\input{nettable}
\end{table}

\FloatBarrier

For additional details please refer to the methods description in~\cite{Lawyer2014}.

\section{Results}

The ExF has consistently high correlation to all spreading process outcomes for all networks tested (mean values SI=0.73, SIS/SIR=0.92), as seen in Figure 1  and detailed in the supplementary tables. 
It strongly and significantly outperformed the other measures, achieving higher measured correlation with tighter confidence bounds. 
The k-shell was the next strongest predictor of outcomes. 
Correlation of the eigenvalue centrality to epidemic outcome showed large variation depending on the network, demonstrating the well known instability of this metric~\cite{Ghoshal2011,Sikic2013}.

\section{Extended background}
\begin{center}
\emph{This section is a summary of key concepts from ~\cite{Lawyer2014}}
\end{center}
The \emph{expected force} was derived to fill a major gap in our understanding of node influence. Most existing measures of influence are designed to produce a ranking which identifies the most important network nodes~\cite{Borgatti2005,Borgatti2006}. 
Key nodes, however, typically comprise less than 1\% of all network nodes, and measures which successfully identify these are often rather less informative for the vast majority of network nodes. Further, a ranking does not quantify the difference between the items ranked.
The ExF fills the gap by accurately quantifying the influence of all network nodes.

The ExF is derived from a continuous-time epidemiological understanding of spread.
Epidemiology defines  the force of infection (FoI) as the current rate at which susceptible nodes are becoming infected~\cite{anderson1992}. 
It is a random variable whose value is determined by the route by which a process spreads in a network. 
For a given node, the expected force is the expected value of the FoI which would arise from spreading processes seeded from that node. 
More precisely, if a node is the start of $w_1$ walks of length one and $w_2$ walks of length two, then the distribution of possible FoI values after two 
transmissions has $O(w_1^2 + w_2)$ terms. The weighting of different walk lengths is determined adoptively by the combinatorics inherent in the local topology surrounding the node.
The definition naturally extends to weighted and/directed networks by including these factors in the calculation of the expected value.

Centrality measures, in contrast, (almost) universally can be expressed as infinite sums of walks~\cite{Estrada2010,Benzi2013}, where the type of walk and decay function in the infinite sum encode the assumption of what makes a node highly influential~\cite{Borgatti2005,Borgatti2006,Estrada2010,Bauer2012,Benzi2013}.
The question not asked is if the type, scaling, and lengths of walks best suited to identifying the most important nodes applies equally well to quantifying the influence of the remaining network nodes.
Given that the best choice depends on the network topology~\cite{Borgatti2006}, the highly heterogeneous topology inherent in complex networks suggests not.

We choose to compare the ExF to the eigenvalue centrality and the k-shell for two reasons. First, these two measures are widely used. Second, if one accepts that  (almost) all centrality measures are based on sums of walks of various lengths~\cite{Borgatti2005,Borgatti2006,Estrada2010, Benzi2013}, then the eigenvalue and the k-shell represent the two endpoints of the distribution of all well-known centrality measures~\cite{Estrada2010, Benzi2013}.

Here, we define node influence in terms of spreading power,  the force with which it can push a spreading process to the rest of the network. 
In a susceptible-infected (SI) process, which inevitably reaches the entire connected component of the network, the spreading power of the seed node predicts the delay before half (or some other large percentage of) the network is reached.
In a process with recovery to either the susceptible (SIS) or immune (SIR) state, spreading power is the probability that a node can seed an epidemic
given that the ratio of the  per-contact transmission rate to the rate of recovery allows for, but does not guarantee, an epidemic.

\vfill

%

\clearpage
\section{Supplement}
The measured correlations used to generate Figure 1.
\begin{table}[h]
 \textbf{Table S1. Correlation between spreading power metrics and time to half coverage in SI processes.}
Shown is the estimated correlation from 1,000 nodes on the given network, along with the 95\% confidence bounds of the estimate.
ExF=expected force, ks=k-shell, evc=eigenvalue centrality.
\begin{ruledtabular}
\begin{tabular}{rcccc} 
 & ExF & ks & evc\\
\hline
2010-10-08 & 0.71 $\pm$ 0.03 & 0.46 $\pm$ 0.05 & 0.39 $\pm$ 0.05 \\
2013-12-01 & 0.75 $\pm$ 0.03 & 0.41 $\pm$ 0.05 & 0.25 $\pm$ 0.06 \\
2014-01-01 & 0.73 $\pm$ 0.03 & 0.44 $\pm$ 0.05 & 0.28 $\pm$ 0.06 \\
2014-02-01 & 0.71 $\pm$ 0.03 & 0.41 $\pm$ 0.05 & 0.32 $\pm$ 0.06 \\
\end{tabular} 
\end{ruledtabular}
\end{table}
\begin{table}[h]
 \textbf{ Table S2. Correlation between spreading power metrics and epidemic potential in discrete time SIS processes.}
Shown is the estimated correlation from 1,000 nodes on the given network, along with the 95\% confidence bounds of the estimate.
ExF=expected force, ks=k-shell, evc=eigenvalue centrality.
\begin{ruledtabular}
\begin{tabular}{rcccc} 
 & ExF & ks & evc\\
\hline
2010-10-08 & 0.93 $\pm$ 0.01 & 0.82 $\pm$ 0.02 & 0.76 $\pm$ 0.03 \\
2013-12-01 & 0.91 $\pm$ 0.01 & 0.80 $\pm$ 0.02 & 0.54 $\pm$ 0.04 \\
2014-01-01 & 0.90 $\pm$ 0.01 & 0.77 $\pm$ 0.03 & 0.55 $\pm$ 0.04 \\
2014-02-01 & 0.93 $\pm$ 0.01 & 0.79 $\pm$ 0.02 & 0.66 $\pm$ 0.03 \\
\end{tabular} 
\end{ruledtabular}
\end{table}
\begin{table}[h]
 \textbf{ Table S3.  Correlation between spreading power metrics and epidemic potential in discrete time SIR processes.}
Shown is the estimated correlation from 1,000 nodes on the given network, along with the 95\% confidence bounds of the estimate. 
ExF=expected force, ks=k-shell, evc=eigenvalue centrality.
\begin{ruledtabular}
\begin{tabular}{rcccc} 
 & ExF & ks & evc\\
\hline
2010-10-08 & 0.93 $\pm$ 0.01 & 0.81 $\pm$ 0.02 & 0.74 $\pm$ 0.03 \\
2013-12-01 & 0.91 $\pm$ 0.01 & 0.80 $\pm$ 0.02 & 0.53 $\pm$ 0.04 \\
2014-01-01 & 0.89 $\pm$ 0.01 & 0.76 $\pm$ 0.03 & 0.54 $\pm$ 0.04 \\
2014-02-01 & 0.93 $\pm$ 0.01 & 0.77 $\pm$ 0.02 & 0.65 $\pm$ 0.04 \\
\end{tabular} 
\end{ruledtabular}
\end{table}

\end{document}

%% file: nettable.tex
\textbf{Table 1. Networks studied here.} Sample networks represent AS level maps of the Internet collected at four separate dates. The networks grow larger with time. This growth slightly shrinks network diameter while dramatically increasing the leading eigenvalue ($\lambda$), though network density remains roughly constant.
\vspace{1ex}

\begin{ruledtabular}
\begin{tabular}{rcccc} 
 date & nodes & diameter & $\lambda$ & density\\
\hline
2010-10-08 & 35,938 & 12 & 283 & 1.60  \\
2013-12-01 & 45,760 & 11 & 422 & 1.56  \\
2014-01-01 & 45,893 & 11 & 413 & 1.54  \\
2014-02-01 & 46,122 & 10 & 417 & 1.56  \\
2014-03-03 & 46,290 & 11 & 419 & 1.57  \\
\end{tabular} 
\end{ruledtabular}

%% file: techreport.bbl
\begin{thebibliography}{12}%
\makeatletter
\providecommand \@ifxundefined [1]{%
 \@ifx{#1\undefined}
}%
\providecommand \@ifnum [1]{%
 \ifnum #1\expandafter \@firstoftwo
 \else \expandafter \@secondoftwo
 \fi
}%
\providecommand \@ifx [1]{%
 \ifx #1\expandafter \@firstoftwo
 \else \expandafter \@secondoftwo
 \fi
}%
\providecommand \natexlab [1]{#1}%
\providecommand \enquote  [1]{``#1''}%
\providecommand \bibnamefont  [1]{#1}%
\providecommand \bibfnamefont [1]{#1}%
\providecommand \citenamefont [1]{#1}%
\providecommand \href@noop [0]{\@secondoftwo}%
\providecommand \href [0]{\begingroup \@sanitize@url \@href}%
\providecommand \@href[1]{\@@startlink{#1}\@@href}%
\providecommand \@@href[1]{\endgroup#1\@@endlink}%
\providecommand \@sanitize@url [0]{\catcode `\\12\catcode `\$12\catcode
  `\&12\catcode `\#12\catcode `\^12\catcode `\_12\catcode `\%12\relax}%
\providecommand \@@startlink[1]{}%
\providecommand \@@endlink[0]{}%
\providecommand \url  [0]{\begingroup\@sanitize@url \@url }%
\providecommand \@url [1]{\endgroup\@href {#1}{\urlprefix }}%
\providecommand \urlprefix  [0]{URL }%
\providecommand \Eprint [0]{\href }%
\providecommand \doibase [0]{http://dx.doi.org/}%
\providecommand \selectlanguage [0]{\@gobble}%
\providecommand \bibinfo  [0]{\@secondoftwo}%
\providecommand \bibfield  [0]{\@secondoftwo}%
\providecommand \translation [1]{[#1]}%
\providecommand \BibitemOpen [0]{}%
\providecommand \bibitemStop [0]{}%
\providecommand \bibitemNoStop [0]{.\EOS\space}%
\providecommand \EOS [0]{\spacefactor3000\relax}%
\providecommand \BibitemShut  [1]{\csname bibitem#1\endcsname}%
\let\auto@bib@innerbib\@empty
\bibitem [{\citenamefont {Estrada}(2010)}]{Estrada2010}%
  \BibitemOpen
  \bibfield  {author} {\bibinfo {author} {\bibfnamefont {E.}~\bibnamefont
  {Estrada}},\ }\href {\doibase 10.1016/j.jtbi.2010.01.014} {\bibfield
  {journal} {\bibinfo  {journal} {J Theor Biol}\ }\textbf {\bibinfo {volume}
  {263}},\ \bibinfo {pages} {556} (\bibinfo {year} {2010})}\BibitemShut
  {NoStop}%
\bibitem [{\citenamefont {Benzi}\ and\ \citenamefont
  {Klymko}(2013)}]{Benzi2013}%
  \BibitemOpen
  \bibfield  {author} {\bibinfo {author} {\bibfnamefont {M.}~\bibnamefont
  {Benzi}}\ and\ \bibinfo {author} {\bibfnamefont {C.}~\bibnamefont {Klymko}},\
  }\href@noop {} {\  (\bibinfo {year} {2013})},\ \Eprint
  {http://arxiv.org/abs/1312.6722} {1312.6722} \BibitemShut {NoStop}%
\bibitem [{\citenamefont {Lawyer}(2014)}]{Lawyer2014}%
  \BibitemOpen
  \bibfield  {author} {\bibinfo {author} {\bibfnamefont {G.}~\bibnamefont
  {Lawyer}},\ }\href {\doibase 10.1038/srep08665} {\bibfield  {journal}
  {\bibinfo  {journal} {Sci. Rep.}\ }\textbf {\bibinfo {volume} {5}},\ 
  \bibinfo{pages}{8665} (\bibinfo {year} {2015})}\BibitemShut {NoStop}%
\bibitem [{\citenamefont {Li}\ \emph {et~al.}(2004)\citenamefont {Li},
  \citenamefont {Alderson}, \citenamefont {Willinger},\ and\ \citenamefont
  {Doyle}}]{Li2004}%
  \BibitemOpen
  \bibfield  {author} {\bibinfo {author} {\bibfnamefont {L.}~\bibnamefont
  {Li}}, \bibinfo {author} {\bibfnamefont {D.}~\bibnamefont {Alderson}},
  \bibinfo {author} {\bibfnamefont {W.}~\bibnamefont {Willinger}}, \ and\
  \bibinfo {author} {\bibfnamefont {J.}~\bibnamefont {Doyle}},\ }\href
  {\doibase 10.1145/1030194.1015470} {\bibfield  {journal} {\bibinfo  {journal}
  {SIGCOMM Comput. Commun. Rev.}\ }\textbf {\bibinfo {volume} {34}},\ \bibinfo
  {pages} {3} (\bibinfo {year} {2004})}\BibitemShut {NoStop}%
\bibitem [{\citenamefont {Pan}\ and\ \citenamefont
  {Saram{\"a}ki}(2011)}]{Pan2011}%
  \BibitemOpen
  \bibfield  {author} {\bibinfo {author} {\bibfnamefont {R.~K.}\ \bibnamefont
  {Pan}}\ and\ \bibinfo {author} {\bibfnamefont {J.}~\bibnamefont
  {Saram{\"a}ki}},\ }\href@noop {} {\bibfield  {journal} {\bibinfo  {journal}
  {Phys. Rev. E}\ }\textbf {\bibinfo {volume} {84}} (\bibinfo {year}
  {2011})}\BibitemShut {NoStop}%
\bibitem [{Note1()}]{Note1}%
  \BibitemOpen
  \bibinfo {note} {Http://irl.cs.ucla.edu/topology/}\BibitemShut {NoStop}%
\bibitem [{\citenamefont {Ghoshal}\ and\ \citenamefont
  {Barabási}(2011)}]{Ghoshal2011}%
  \BibitemOpen
  \bibfield  {author} {\bibinfo {author} {\bibfnamefont {G.}~\bibnamefont
  {Ghoshal}}\ and\ \bibinfo {author} {\bibfnamefont {A.~L.}\ \bibnamefont
  {Barabási}},\ }\href {\doibase 10.1038/ncomms1396} {\bibfield  {journal}
  {\bibinfo  {journal} {Nat Commun}\ }\textbf {\bibinfo {volume} {2}},\
  \bibinfo {pages} {394} (\bibinfo {year} {2011})}\BibitemShut {NoStop}%
\bibitem [{\citenamefont {Sikic}\ \emph {et~al.}(2013)\citenamefont {Sikic},
  \citenamefont {Lancic}, \citenamefont {Antulov-Fantulin},\ and\ \citenamefont
  {Stefancic}}]{Sikic2013}%
  \BibitemOpen
  \bibfield  {author} {\bibinfo {author} {\bibfnamefont {M.}~\bibnamefont
  {Sikic}}, \bibinfo {author} {\bibfnamefont {A.}~\bibnamefont {Lancic}},
  \bibinfo {author} {\bibfnamefont {N.}~\bibnamefont {Antulov-Fantulin}}, \
  and\ \bibinfo {author} {\bibfnamefont {H.}~\bibnamefont {Stefancic}},\ }\href
  {\doibase 10.1140/epjb/e2013-31025-5} {\bibfield  {journal} {\bibinfo
  {journal} {The European Physical Journal B}\ }\textbf {\bibinfo {volume}
  {86}},\ \bibinfo {pages} {1} (\bibinfo {year} {2013})}\BibitemShut {NoStop}%
\bibitem [{\citenamefont {Borgatti}(2005)}]{Borgatti2005}%
  \BibitemOpen
  \bibfield  {author} {\bibinfo {author} {\bibfnamefont {S.~P.}\ \bibnamefont
  {Borgatti}},\ }\href {\doibase 10.1016/j.socnet.2004.11.008} {\bibfield
  {journal} {\bibinfo  {journal} {Social Networks}\ }\textbf {\bibinfo {volume}
  {27}},\ \bibinfo {pages} {55} (\bibinfo {year} {2005})}\BibitemShut {NoStop}%
\bibitem [{\citenamefont {Borgatti}\ and\ \citenamefont
  {Everett}(2006)}]{Borgatti2006}%
  \BibitemOpen
  \bibfield  {author} {\bibinfo {author} {\bibfnamefont {S.~P.}\ \bibnamefont
  {Borgatti}}\ and\ \bibinfo {author} {\bibfnamefont {M.~G.}\ \bibnamefont
  {Everett}},\ }\href@noop {} {\bibfield  {journal} {\bibinfo  {journal}
  {Social Networks}\ }\textbf {\bibinfo {volume} {28}},\ \bibinfo {pages} {466}
  (\bibinfo {year} {2006})}\BibitemShut {NoStop}%
\bibitem [{\citenamefont {Anderson}\ and\ \citenamefont
  {May}(1992)}]{anderson1992}%
  \BibitemOpen
  \bibfield  {author} {\bibinfo {author} {\bibfnamefont {R.~M.}\ \bibnamefont
  {Anderson}}\ and\ \bibinfo {author} {\bibfnamefont {R.~M.}\ \bibnamefont
  {May}},\ }\href@noop {} {\emph {\bibinfo {title} {Infectious Diseases of
  Humans: Dynamics and Control}}}\ (\bibinfo  {publisher} {Oxford University
  Press},\ \bibinfo {year} {1992})\BibitemShut {NoStop}%
\bibitem [{\citenamefont {Bauer}\ and\ \citenamefont
  {Lizier}(2012)}]{Bauer2012}%
  \BibitemOpen
  \bibfield  {author} {\bibinfo {author} {\bibfnamefont {F.}~\bibnamefont
  {Bauer}}\ and\ \bibinfo {author} {\bibfnamefont {J.~T.}\ \bibnamefont
  {Lizier}},\ }\href {http://stacks.iop.org/0295-5075/99/i=6/a=68007}
  {\bibfield  {journal} {\bibinfo  {journal} {Europhys Lett}\ }\textbf
  {\bibinfo {volume} {99}},\ \bibinfo {pages} {68007} (\bibinfo {year}
  {2012})}\BibitemShut {NoStop}%
\end{thebibliography}
